\documentclass{sf2a-conf2015}
\usepackage{txfonts}
\usepackage{graphicx}
\usepackage{hyperref}
\usepackage[]{natbib}  
\usepackage{epstopdf}

\def\BibTeX{{\rm B\kern-.05em{\sc i\kern-.025em b}\kern-.08em
    T\kern-.1667em\lower.7ex\hbox{E}\kern-.125emX}}
\bibpunct{(}{)}{;}{a}{}{,}  

\begin{document}

\TitreGlobal{SF2A 2015}


\title{The variation of the tidal quality factor of convective envelopes\\ of rotating low-mass stars along their evolution}

\runningtitle{The variation of the tidal quality factor of convective envelopes of rotating low-mass stars}

\author{S. Mathis}\address{Laboratoire AIM Paris-Saclay, CEA/DSM - CNRS - Universit\'e Paris Diderot, IRFU/SAp Centre de Saclay, F-91191 Gif-sur-Yvette Cedex, France}

\setcounter{page}{237}


\maketitle


\begin{abstract}
More than 1500 exoplanets have been discovered around a large diversity of host stars (from M- to A-type stars). Tidal dissipation in their convective envelope is a key actor that shapes the orbital architecture of short-period systems and that still remains unknown. Using a simplified two-layer assumption and grids of stellar models, we compute analytically an equivalent modified tidal quality factor, which is proportional to the inverse of the frequency-averaged dissipation due to the viscous friction applied by turbulent convection on tidal waves. It leads the conversion of their kinetic energy into heat and tidal evolution of orbits and spin. During their Pre-Main-Sequence, all low-mass stars have a decrease of the equivalent modified tidal quality factor for a fixed angular velocity of their convective envelope. Next, it evolves on the Main Sequence to an asymptotic value that is minimum for $0.6M_{\odot}$ K-type stars and that increases by several orders of magnitude with increasing stellar mass. Finally, the rotational evolution of low-mass stars strengthens tidal dissipation during the Pre-Main-Sequence.
\end{abstract}



\section{Introduction and context}

Since twenty years, more than 1500 exoplanets have been discovered \citep[e.g.][]{Perryman2011}. The orbital architecture of their systems is strongly different from the one we know for our solar system \cite[e.g.][]{Fabetal2014}. It stimulates a lot of studies of their dynamical evolution and stability \citep[e.g.][]{Bolmontetal2012,Laskaretal2012}. In such studies, tidal interactions are one of the principal physical mechanisms that must be modeled, particularly for short-period systems. Indeed, the dissipation of the kinetic energy of tidal flows/displacements in stellar and planetary interiors leads to a modification of the semi-major axis and the eccentricity of the orbits, of the rotation of celestial bodies and of the relative inclination of the spins \citep[][]{Hut1980,Hut1981}. In this framework, a large majority of works proposed to use the so-called {\it tidal quality factor} $Q$ to parametrize this dissipation and the related friction processes \citep{Kaula1964}. As in the case of the theory of forced damped oscillators, the dissipation is strong and the evolution is rapid when the quality factor is small and vice-versa. Two ways are then proposed to choose a value for $Q$: i) one can choose to calibrate it on observations or on formation scenario \citep[][]{GoldreichSoter1966,Hansen2012}; ii) one choose to compute it using an ab-initio treatment of dissipative mechanisms acting on tidal motions \citep[e.g.][]{MathisRemus2013,Ogilvie2014,LeBarsetal2015}. In the second case, it is now demonstrated that tidal dissipation is a complex function of the internal structure of celestial bodies, of their dynamical properties (their rotation, stratification, viscosity, thermal diffusivity, etc.) and of the forcing frequency \citep[e.g.][and references therein]{EL2007,MathisRemus2013,Ogilvie2014,ADMLP2015}. Such dependences have a strong impact on the dynamical evolution of systems \citep[][]{ADLPM2014}.\\ 

To obtain a coherent picture of the dynamics of exoplanetary systems it is thus necessary to have a correct evaluation of tidal dissipation in their host stars along their evolution; for instance this dissipation has a strong impact on the orbital configuration of short-period systems. From now on, stellar mass range spreads from M red dwarfs to intermediate-mass A-type stars. In this context, tidal friction in the rotating turbulent convective envelopes of these low-mass stars plays an important role for tidal migration,  circularization of orbits,  synchronization and alignment of spins \citep[e.g.][and references therein for hot-Jupiter systems]{Winnetal2010,Albrechtetal2012,Lai2012,Ogilvie2014,VR2014}. In stellar convective layers, tidal flows are constituted of large-scale non-wavelike/equilibrium flows driven by the adjustment of the hydrostatic structure of stars because of the presence of the planetary/stellar companion \citep{Zahn1966b,RMZ2012} and the dynamical tide constituted by inertial waves, which have the Coriolis acceleration as restoring force \cite[e.g.][]{OL2007}. In this framework, both the structure and rotation of stars strongly varies along their evolution \cite[e.g.][]{Siessetal2000,GB2013,GB2015}. Moreover, as reported by \cite{Ogilvie2014}, observations of star-planet and binary-star systems show that tidal dissipation varies over several orders of magnitude. Therefore, the key questions that must be addressed for dynamical studies is {\it how does the tidal quality factor of the convective envelope of low-mass stars vary as a function of stellar mass, evolutionary stage, and rotation?}

Using results presented in \cite{Mathis2015}, we introduce here an equivalent modified tidal quality factor, which has been defined by \cite{OL2007} and is proportional to the inverse of the frequency-averaged dissipation. We compute it as a function of the mass, the age, and the rotation of stars using realistic computed grids of stellar models which can be used for dynamical studies of star-planet systems in a first step. In sec. 2, we introduce the assumptions and the formalism that allows us to analytically evaluate this quantity as a function of the structure and rotation of stars \citep{OL2007,Ogilvie2013}. In sec. 3, we compute it as a function of stellar mass and evolutionary stage at fixed angular velocity using grids of stellar models for low-mass stars from $0.4$ to $1.4M_{\odot}$. In sec. 4, we present our conclusions.

\section{Tidal dissipation modelling}

\begin{figure}[!h]
\centering
\includegraphics[width=0.38\textwidth]{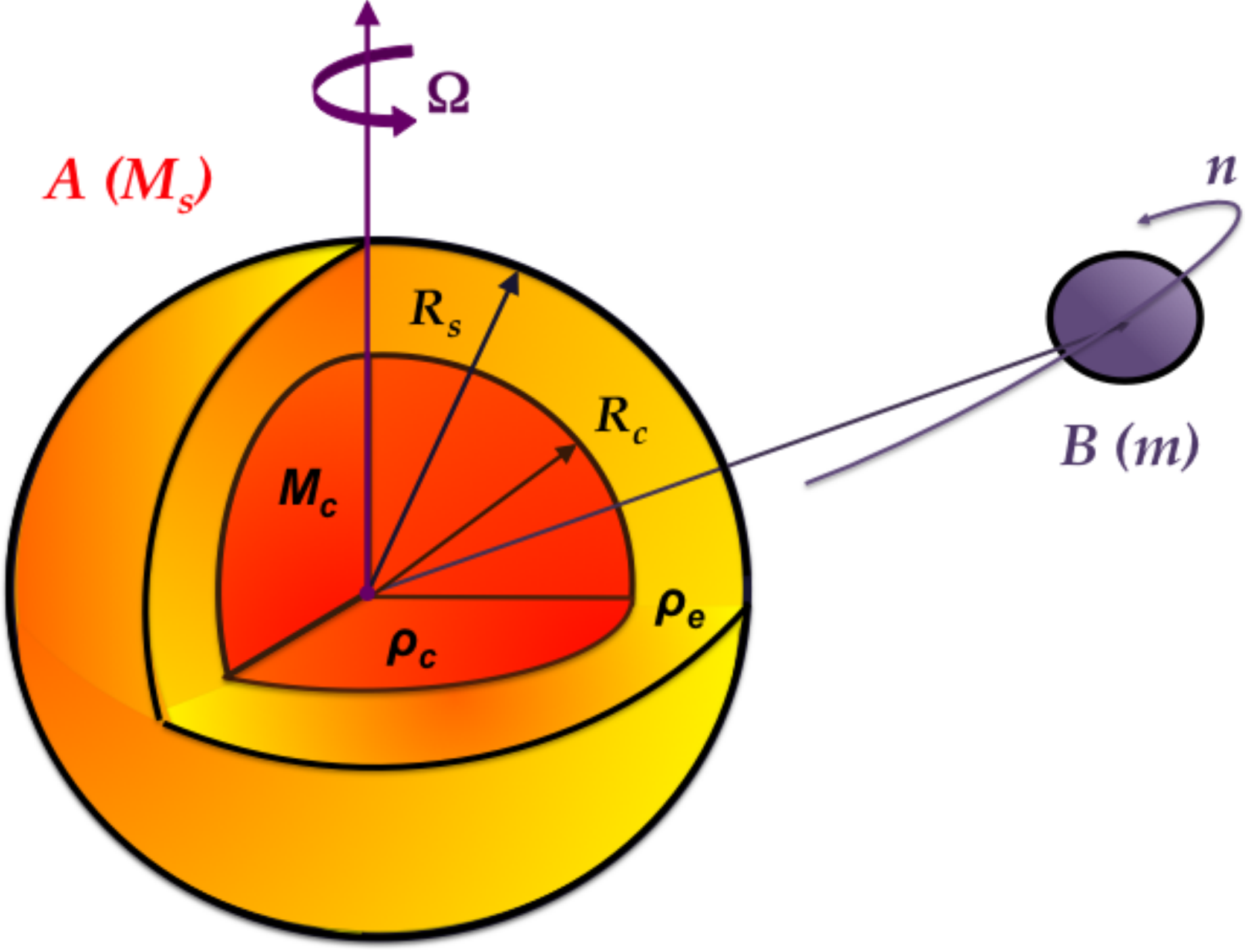}
\caption{Two-layer low-mass star A of mass $M_{\rm s}$ and mean radius $R_{\rm s}$ and point-mass tidal perturber B of mass $m$ orbiting with a mean motion $n$. The radiative core of radius $R_{\rm c}$, mass $M_{\rm c}$, and density $\rho_{\rm c}$ is surrounded by the convective envelope of density $\rho_{\rm e}$.}
\label{SetUp}
\end{figure}

In this work, we adopt a simplified two-layer model of a star A of mass $M_{\rm s}$ and mean radius $R_{\rm s}$ hosting a point-mass tidal perturber B of mass $m$ orbiting with a mean motion $n$ \citep[e.g.][and Fig. \ref{SetUp}]{Ogilvie2013,PZJ2014}. In this model, both the central radiative region and the convective envelope are assumed to be homogeneous with respective constant densities $\rho_{\rm c}$ and $\rho_{\rm e}$. The convective layers of A are assumed to be in a moderate solid-body rotation with an angular velocity $\Omega$ so that $\epsilon^2 \equiv \left(\Omega/ \sqrt{\mathcal{G} M_{\rm s} / R_{\rm s}^3}\right)^2=\left(\Omega/\Omega_{\rm c}\right)^2 \ll 1$, where $\Omega_{\rm c}$ {is the critical angular velocity} and ${\mathcal G}$ is the gravitational constant. Therefore, the centrifugal acceleration, which scales as $\Omega^{2}$, is not taken into account. It surrounds the radiative core of radius $R_{\rm c}$ and mass $M_{\rm c}$. 

Tidal dissipation in the convection zone of A originates from the excitation by B of inertial waves, which have the Coriolis acceleration as restoring force and are excited if the tidal forcing frequency $\omega\in\left[-2\Omega,2\Omega\right]$. They are damped by the turbulent friction, which is modelled using a turbulent viscosity \citep[see e.g.][and references therein]{OL2012}. Its analytical evaluation in our two-layer model was conducted by \cite{Ogilvie2013} who assumed an incompressible convective envelope, which corresponds to inertial waves with shorter wavelength than the characteristic length of variation of the density. In this modeling, the dissipation of tidal internal gravity waves in the stable radiative core is not taken into account \citep[][]{BO2010,Ivanovetal2013} and the higher-frequency acoustic waves are filtered out. It allows us to compute the frequency-averaged tidal dissipation {given in \cite{Ogilvie2013} (Eq. B3)} and to introduce an equivalent modified tidal quality factor, ${\overline {Q'}}$, as defined by \cite{OL2007}\footnote{We point out here that equivalent quality factors ${\overline{Q'}}$ and ${\overline Q}$, which are proportional to the inverse of the frequency-averaged dissipation $\left<{\rm Im} \left[k_2^2(\omega)\right]\right>_{\omega}$, where $\left<...\right>_{\omega}=\int_{-\infty}^{\infty}...{\mathrm{d}\omega}/{\omega}$, are not equivalent to potentially defined frequency-averaged quality factors $\left<Q'\left(\omega\right)\right>_{\omega}$ and $\left<Q\left(\omega\right)\right>_{\omega}$. In this framework, the relevant physical quantity being $\left<{\rm Im} \left[k_2^2(\omega)\right]\right>_{\omega}$, we prefer to define directly equivalent quality factors from it.}:
\begin{eqnarray}
\lefteqn{\frac{3}{2{\overline {Q'}}}=\frac{k_2}{{\overline Q}}=\int^{+\infty}_{-\infty} \! {\rm Im} \left[k_2^2(\omega)\right] \,\frac{\mathrm{d}\omega}{\omega}=\left<{\rm Im} \left[k_2^2(\omega)\right]\right>_{\omega} = \frac{100 \pi}{63} \epsilon^2 \left(\frac{\alpha^5}{1-\alpha^5}\right)\left(1-\gamma\right)^2}\nonumber\\
&&\times\left(1-\alpha\right)^4\left(1+2\alpha+3\alpha^2+\frac{3}{2}\alpha^3\right)^2\left[1+\left(\frac{1-\gamma}{\gamma}\right)\alpha^3\right]\left[1+\frac{3}{2}\gamma+\frac{5}{2\gamma}\left(1+\frac{1}{2}\gamma-\frac{3}{2}\gamma^2\right)\alpha^3-\frac{9}{4}\left(1-\delta\right)\alpha^5\right]^{-2}
\label{eq:imk22fogilvie}
\end{eqnarray}
with
\begin{equation}
\alpha=\frac{R_{\rm c}}{R_{\rm s}}\hbox{,}\quad\beta=\frac{M_{\rm c}}{M_{\rm s}}\quad\hbox{and}\quad\gamma=\frac{\rho_{\rm e}}{\rho_{\rm c}}=\frac{\alpha^3\left(1-\beta\right)}{\beta\left(1-\alpha^3\right)}<1.
\end{equation}
We introduce the quadrupolar complex Love number $k_{2}^{2}$, associated with the $\left(2,2\right)$ component of the time-dependent tidal potential $U$ that corresponds to the spherical harmonic $Y_2^2$. It quantifies at the surface of the star ($r=R_{\rm s}$) the ratio of the tidal perturbation of its self-gravity potential over the tidal potential in the simplest case of coplanar systems. In the case of dissipative convective envelopes, it is a complex quantity which depends on the tidal frequency ($\omega$) with a real part that accounts for the energy stored in the tidal perturbation while the imaginary part accounts for the energy losses \citep[e.g.][]{RMZ2012}. We also recall the correspondance with the commonly used real hydrostatic quadrupolar Love number $k_{2}$, which is independent of $\omega$ in the case of fluid layers, and an equivalent tidal quality factor ${\overline Q}$. The dissipation being averaged in frequency (with the corresponding notation $\left<...\right>_{\omega}$), its complicated frequency-dependence in a spherical shell \citep[][]{OL2007} is filtered out. The dissipation at a given frequency could thus be larger or smaller than its averaged value by several orders of magnitude.

\section{Tidal dissipation along stellar evolution and the variation of ${\overline{Q'}}$}

\begin{figure*}[!t]
\begin{center}
\includegraphics[width=0.775\linewidth]{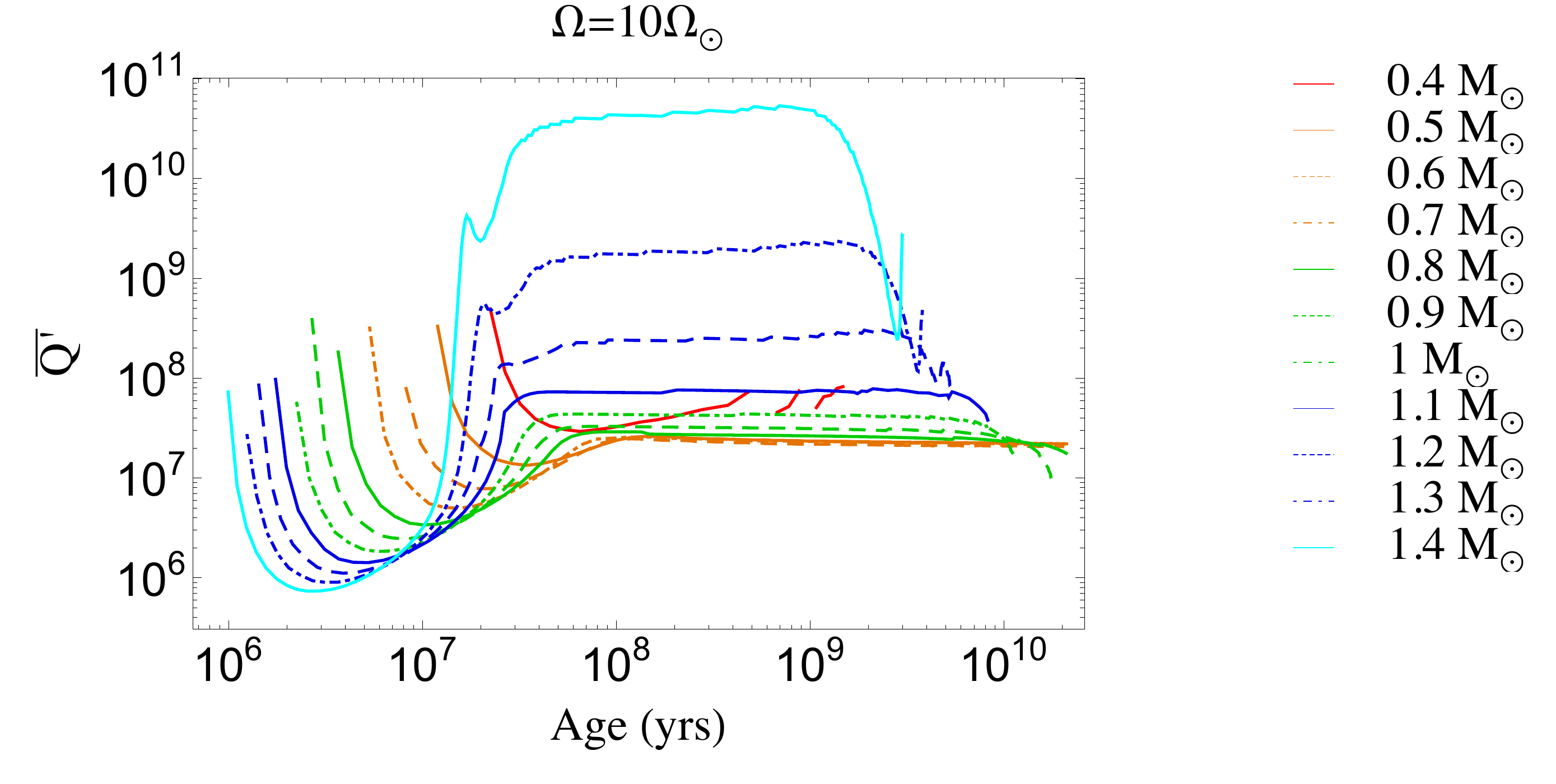}
\end{center} 
\caption{Evolution of the frequency-averaged modified tidal quality factor, ${\overline{Q'}}$, for $\Omega=10\Omega_{\odot}$, where $\Omega_{\odot}$ is the solar angular velocity, as a function of time for stellar masses ($M_{\rm s}$) from $0.4$ to $1.4M_{\odot}$.}
\label{TidalAmplitude}
\end{figure*}

Following \cite{Mathis2015}, we compute the variation of the equivalent modified tidal quality factor ${\overline{Q'}}$ (Eq. \ref{eq:imk22fogilvie}) as a function of stellar mass, age and rotation. To reach this objective, we compute the radius and mass aspect ratios (respectively $\alpha$ and $\beta$) as function of time using grids of stellar models from $0.4$ to $1.4M_{\odot}$ for a metallicity $Z=0.02$ computed by \cite{Siessetal2000} using the STAREVOL code. It allows us to plot in Fig. \ref{TidalAmplitude} the variation of ${\overline{Q'}}$ with time with taking into account the simultaneous variations of $\alpha$, $\beta$, and $R_{\rm s}$. On the Pre-Main-Sequence (hereafter PMS), it decreases towards the minimum value, which decreases with stellar mass, corresponding to the region around $\left(\alpha_{\rm max}\approx0.571,\beta_{\rm max}\approx0.501\right)$ where the dissipation is maximum for all stellar masses. This is due to the formation and the growth of the radiative core that leads to spherical shell configurations where dissipative inertial wave attractors may take place \citep{O2005,GL2009}. The time coordinate of this minimum is smaller if stellar mass is higher because of the corresponding shorter life-time of the star. Then, ${\overline{Q'}}$ increases to rapidly reach its almost constant value on the Main Sequence (hereafter MS). As already pointed out,  and expected from observational constraints \citep[e.g.][]{Albrechtetal2012} and previous theoretical works \citep{OL2007,BO2009}, it increases with stellar mass from {\bf $0.6M_{\odot}$} K-type to A-type stars by several orders of magnitude ({$\approx3$ between $0.6$} and $1.4M_{\odot}$) because of the variation in the thickness of the convective envelope, which becomes thinner \footnote{In the cases of F- and A-type stars, a convective core is present in addition to the envelope. We neglect its action assuming that it corresponds to the regime of weak dissipation in a full sphere \citep[][]{Wu2005}.}. For $0.4M_{\odot}$ M-type stars, the radiative core has an evolution where it finally disappears. Therefore, the configuration converges on the MS to the case of  fully convective stars with the weak dissipation of normal inertial modes \citep{Wu2005}. Finally, for higher mass stars, we can see a final decrease of ${\overline{Q'}}$ because of the simultaneous extension of their convective envelope and contraction of their radiative core during their Sub Giant phase leading again the star towards the region of maximum dissipation in the $(\alpha,\beta)$ plane.\\

It is also crucial to discuss consequences of the rotational evolution of low-mass stars. As we know from observational works and related modelling \citep[e.g.][and references therein]{GB2013,GB2015}, their rotation follows three phases of evolution: i)  stars are trapped in co-rotation with the surrounding circumstellar disk; ii)  because of the contraction of stars on the PMS their rotation (and thus $\epsilon$) increases; and iii)  stars are braked on the MS because of the torque applied by pressure-driven stellar winds \citep[e.g.][and references therein]{Revilleetal2015} and $\epsilon$ decreases. As a detailed computation of stellar rotating models is beyond the scope of the present work, we have here computed ${\overline{Q'}}$ for an intermediate rotation $\Omega=10\Omega_{\odot}$ (the values for $\Omega=\Omega_{\odot}$ and $\Omega=100\Omega_{\odot}$ will be 100 times higher and lower respectively). From results obtained by \cite{GB2013,GB2015}, we can easily infer that ${\overline{Q'}}$ decreases during the PMS because of the growth of the radiative core and of the angular velocity. On the MS, ${\overline{Q'}}$ increases because of the evolution of the structure of the star and of its braking by stellar winds.

\section{Conclusions}

All low-mass stars have a decrease of the equivalent modified tidal quality factor for a fixed angular velocity in their convective envelope for tidal frequencies lying within the range $\left[-2\Omega,2\Omega\right]$ so that inertial waves can be excited until they reach a critical aspect and mass ratios close to ($\alpha_{\rm max},\beta_{\rm max}$) during the PMS. Next, it evolves on the MS to an asymptotic value that {reaches a} minimum for $0.6M_{\odot}$ K-type stars and then increases by several orders of magnitude with increasing stellar mass. Finally, {the rotational evolution of low-mass stars strengthens the importance of tidal dissipation during the PMS as pointed out  by \cite{Zahn1989}}.

In the near future, it would be important to relax current assumptions by taking into account density stratification, differential rotation, magnetic field, non-linear effects (and related possible instabilities) and  the dissipation of tidal gravito-inertial waves in stellar radiative cores \citep[e.g.][]{BR2013,Schmitt2010,Favieretal2014,Ivanovetal2013} to get a complete picture \citep{Guillotetal2014}. Finally, it will be interesting to explore advanced phases of stellar evolution.

\begin{acknowledgements}
This work was supported by the Programme National de Plan\'etologie (CNRS/INSU) and the CoRoT-CNES grant at Service d'Astrophysique (CEA-Saclay).
\end{acknowledgements}

\bibliographystyle{aa}  
\bibliography{Mathis} 

\begin{thebibliography}{41}
\expandafter\ifx\csname natexlab\endcsname\relax\def\natexlab#1{#1}\fi

\bibitem[{{Albrecht} {et~al.}(2012){Albrecht}, {Winn}, {Johnson}, {Howard},
  {Marcy}, {Butler}, {Arriagada}, {Crane}, {Shectman}, {Thompson}, {Hirano},
  {Bakos}, \& {Hartman}}]{Albrechtetal2012}
{Albrecht}, S., {Winn}, J.~N., {Johnson}, J.~A., {et~al.} 2012, \apj, 757, 18

\bibitem[{{Auclair-Desrotour} {et~al.}(2014){Auclair-Desrotour}, {Le
  Poncin-Lafitte}, \& {Mathis}}]{ADLPM2014}
{Auclair-Desrotour}, P., {Le Poncin-Lafitte}, C., \& {Mathis}, S. 2014, \aap,
  561, L7

\bibitem[{{Auclair Desrotour} {et~al.}(2015){Auclair Desrotour}, {Mathis}, \&
  {Le Poncin-Lafitte}}]{ADMLP2015}
{Auclair Desrotour}, P., {Mathis}, S., \& {Le Poncin-Lafitte}, C. 2015, \aap,
  581, A118

\bibitem[{{Barker} \& {Ogilvie}(2009)}]{BO2009}
{Barker}, A.~J. \& {Ogilvie}, G.~I. 2009, \mnras, 395, 2268

\bibitem[{{Barker} \& {Ogilvie}(2010)}]{BO2010}
{Barker}, A.~J. \& {Ogilvie}, G.~I. 2010, \mnras, 404, 1849

\bibitem[{{Baruteau} \& {Rieutord}(2013)}]{BR2013}
{Baruteau}, C. \& {Rieutord}, M. 2013, Journal of Fluid Mechanics, 719, 47

\bibitem[{{Bolmont} {et~al.}(2012){Bolmont}, {Raymond}, {Leconte}, \&
  {Matt}}]{Bolmontetal2012}
{Bolmont}, E., {Raymond}, S.~N., {Leconte}, J., \& {Matt}, S.~P. 2012, \aap,
  544, A124

\bibitem[{{Efroimsky} \& {Lainey}(2007)}]{EL2007}
{Efroimsky}, M. \& {Lainey}, V. 2007, Journal of Geophysical Research
  (Planets), 112, 12003

\bibitem[{{Fabrycky} {et~al.}(2014){Fabrycky}, {Lissauer}, {Ragozzine}, {Rowe},
  {Steffen}, {Agol}, {Barclay}, {Batalha}, {Borucki}, {Ciardi}, {Ford},
  {Gautier}, {Geary}, {Holman}, {Jenkins}, {Li}, {Morehead}, {Morris},
  {Shporer}, {Smith}, {Still}, \& {Van Cleve}}]{Fabetal2014}
{Fabrycky}, D.~C., {Lissauer}, J.~J., {Ragozzine}, D., {et~al.} 2014, \apj,
  790, 146

\bibitem[{{Favier} {et~al.}(2014){Favier}, {Barker}, {Baruteau}, \&
  {Ogilvie}}]{Favieretal2014}
{Favier}, B., {Barker}, A.~J., {Baruteau}, C., \& {Ogilvie}, G.~I. 2014,
  \mnras, 439, 845

\bibitem[{{Gallet} \& {Bouvier}(2013)}]{GB2013}
{Gallet}, F. \& {Bouvier}, J. 2013, \aap, 556, A36

\bibitem[{{Gallet} \& {Bouvier}(2015)}]{GB2015}
{Gallet}, F. \& {Bouvier}, J. 2015, ArXiv e-prints

\bibitem[{{Goldreich} \& {Soter}(1966)}]{GoldreichSoter1966}
{Goldreich}, P. \& {Soter}, S. 1966, \icarus, 5, 375

\bibitem[{{Goodman} \& {Lackner}(2009)}]{GL2009}
{Goodman}, J. \& {Lackner}, C. 2009, \apj, 696, 2054

\bibitem[{{Guillot} {et~al.}(2014){Guillot}, {Lin}, {Morel}, {Havel}, \&
  {Parmentier}}]{Guillotetal2014}
{Guillot}, T., {Lin}, D.~N.~C., {Morel}, P., {Havel}, M., \& {Parmentier}, V.
  2014, in EAS Publications Series, Vol.~65, EAS Publications Series, 327--336

\bibitem[{{Hansen}(2012)}]{Hansen2012}
{Hansen}, B.~M.~S. 2012, \apj, 757, 6

\bibitem[{{Hut}(1980)}]{Hut1980}
{Hut}, P. 1980, \aap, 92, 167

\bibitem[{{Hut}(1981)}]{Hut1981}
{Hut}, P. 1981, \aap, 99, 126

\bibitem[{{Ivanov} {et~al.}(2013){Ivanov}, {Papaloizou}, \&
  {Chernov}}]{Ivanovetal2013}
{Ivanov}, P.~B., {Papaloizou}, J.~C.~B., \& {Chernov}, S.~V. 2013, \mnras, 432,
  2339

\bibitem[{{Kaula}(1964)}]{Kaula1964}
{Kaula}, W.~M. 1964, Reviews of Geophysics and Space Physics, 2, 661

\bibitem[{{Lai}(2012)}]{Lai2012}
{Lai}, D. 2012, \mnras, 423, 486

\bibitem[{{Laskar} {et~al.}(2012){Laskar}, {Bou{\'e}}, \&
  {Correia}}]{Laskaretal2012}
{Laskar}, J., {Bou{\'e}}, G., \& {Correia}, A.~C.~M. 2012, \aap, 538, A105

\bibitem[{{Le Bars} {et~al.}(2015){Le Bars}, {C{\'e}bron}, \& {Le
  Gal}}]{LeBarsetal2015}
{Le Bars}, M., {C{\'e}bron}, D., \& {Le Gal}, P. 2015, Annual Review of Fluid
  Mechanics, 47, 163

\bibitem[{{Mathis}(2015)}]{Mathis2015}
{Mathis}, S. 2015, \aap, 580, L3

\bibitem[{{Mathis} \& {Remus}(2013)}]{MathisRemus2013}
{Mathis}, S. \& {Remus}, F. 2013, in Lecture Notes in Physics, Berlin Springer
  Verlag, Vol. 857, Lecture Notes in Physics, Berlin Springer Verlag, ed. J.-P.
  {Rozelot} \& C.~. {Neiner}, 111--147

\bibitem[{{Ogilvie}(2005)}]{O2005}
{Ogilvie}, G.~I. 2005, Journal of Fluid Mechanics, 543, 19

\bibitem[{{Ogilvie}(2013)}]{Ogilvie2013}
{Ogilvie}, G.~I. 2013, \mnras, 429, 613

\bibitem[{{Ogilvie}(2014)}]{Ogilvie2014}
{Ogilvie}, G.~I. 2014, \araa, 52, 171

\bibitem[{{Ogilvie} \& {Lesur}(2012)}]{OL2012}
{Ogilvie}, G.~I. \& {Lesur}, G. 2012, \mnras, 422, 1975

\bibitem[{{Ogilvie} \& {Lin}(2007)}]{OL2007}
{Ogilvie}, G.~I. \& {Lin}, D.~N.~C. 2007, \apj, 661, 1180

\bibitem[{{Penev} {et~al.}(2014){Penev}, {Zhang}, \& {Jackson}}]{PZJ2014}
{Penev}, K., {Zhang}, M., \& {Jackson}, B. 2014, \pasp, 126, 553

\bibitem[{{Perryman}(2011)}]{Perryman2011}
{Perryman}, M. 2011, {The Exoplanet Handbook}

\bibitem[{{Remus} {et~al.}(2012){Remus}, {Mathis}, \& {Zahn}}]{RMZ2012}
{Remus}, F., {Mathis}, S., \& {Zahn}, J.-P. 2012, \aap, 544, A132

\bibitem[{{R{\'e}ville} {et~al.}(2015){R{\'e}ville}, {Brun}, {Matt},
  {Strugarek}, \& {Pinto}}]{Revilleetal2015}
{R{\'e}ville}, V., {Brun}, A.~S., {Matt}, S.~P., {Strugarek}, A., \& {Pinto},
  R.~F. 2015, \apj, 798, 116

\bibitem[{{Schmitt}(2010)}]{Schmitt2010}
{Schmitt}, D. 2010, Geophysical and Astrophysical Fluid Dynamics, 104, 135

\bibitem[{{Siess} {et~al.}(2000){Siess}, {Dufour}, \&
  {Forestini}}]{Siessetal2000}
{Siess}, L., {Dufour}, E., \& {Forestini}, M. 2000, \aap, 358, 593

\bibitem[{{Valsecchi} \& {Rasio}(2014)}]{VR2014}
{Valsecchi}, F. \& {Rasio}, F.~A. 2014, \apj, 786, 102

\bibitem[{{Winn} {et~al.}(2010){Winn}, {Fabrycky}, {Albrecht}, \&
  {Johnson}}]{Winnetal2010}
{Winn}, J.~N., {Fabrycky}, D., {Albrecht}, S., \& {Johnson}, J.~A. 2010, \apjl,
  718, L145

\bibitem[{{Wu}(2005)}]{Wu2005}
{Wu}, Y. 2005, \apj, 635, 688

\bibitem[{{Zahn}(1966)}]{Zahn1966b}
{Zahn}, J.~P. 1966, Annales d'Astrophysique, 29, 489

\bibitem[{{Zahn} \& {Bouchet}(1989)}]{Zahn1989}
{Zahn}, J.-P. \& {Bouchet}, L. 1989, \aap, 223, 112

\end{thebibliography}

\end{document}